\documentclass[aps,pra,showpacs,superscriptaddress,twocolumn,
amsmath,amssymb,longbibliography]{revtex4-1}
\usepackage{amsfonts}
\usepackage{txfonts}
\usepackage{amsmath}
\usepackage{float}
\usepackage[colorlinks,breaklinks,linkcolor=blue,anchorcolor=blue,citecolor=blue,urlcolor=blue
]{hyperref}

\usepackage{graphicx}
\usepackage{dcolumn}
\usepackage{bm}
\usepackage{amssymb}
\usepackage{mathrsfs}
\usepackage[sort&compress]{natbib}
\usepackage{subfigure}
 
\begin{document}
\title{Quantum $\varphi$-synchronization in coupled optomechanical system with periodic modulation}
\author{G. J. Qiao}
\affiliation{Center for Quantum Sciences and School of Physics, Northeast Normal University, Changchun 130024, China}
\author{X. Y. Liu}
\affiliation{Center for Quantum Sciences and School of Physics, Northeast Normal University, Changchun 130024, China}
\author{H. D. Liu}
\email[]{liuhd100@nenu.edu.cn}
\affiliation{Center for Quantum Sciences and School of Physics, Northeast Normal University, Changchun 130024, China}
\affiliation{Department of Physics, University of Texas at Austin, Austin, Texas 78712, USA}
\author{C. F. Sun}
\email[]{suncf997@nenu.edu.cn}
\affiliation{Center for Quantum Sciences and School of Physics, Northeast Normal University, Changchun 130024, China}
\author{X. X. Yi}
\affiliation{Center for Quantum Sciences and School of Physics, Northeast Normal University, Changchun 130024, China}
\date{\today}

\begin{abstract}
Based on the concepts of quantum synchronization and quantum phase synchronization proposed by A. Mari \textit{et al.} in Phys. Rev. Lett. 111, 103605 (2013), we introduce and characterize the measure of a more generalized quantum synchronization called quantum $\varphi$-synchronization  under which the pairs of variables have the same amplitude and possess the same $\varphi$ phase shift.  Naturally, quantum synchronization and quantum anti-synchronization become special cases of quantum $\varphi$-synchronization. Their relations and differences are also discussed. To illustrate these theories, we investigate the quantum $\varphi$-synchronization and quantum phase synchronization phenomena of two coupled optomechanical systems with periodic modulation and show that quantum $\varphi$-synchronization is more general as a measure of synchronization. We also show the phenomenon of quantum anti-synchronization when $\varphi=\pi$.
\end{abstract}
\maketitle

\section{Introduction}
As a collective dynamic behavior in complex systems, synchronization was first proposed by Huygens in the 17th century   \cite{10.1143/PTP.69.32,PhysRevLett.89.054101}. He noticed that the oscillations of two pendulum clocks with a common support tend to synchronize with each other \cite{huygens1897oeuvres}. Since then, synchronization has been widely studied and applied in classical physics. Furthermore, with the development of quantum mechanics, the concept of quantum synchronization was proposed and widely applied in the fields, such as cavity quantum
electrodynamics \cite{PhysRevB.80.014519,PhysRevA.91.012301}, atomic ensembles \cite{PhysRevLett.113.154101,PhysRevLett.114.103601,PhysRevA.91.061401}, van der Pol
(VdP) oscillators \cite{PhysRevA.91.012301,PhysRevE.89.022913,PhysRevLett.112.094102,Eshaqi-Sani2019}, Bose-Einstein condensation \cite{samoylova2015synchronization},
superconducting circuit systems \cite{gul2014synchronization,PhysRevLett.111.073602}, and so on.

In recent years, there has been a growing interest in exploiting synchronization \cite{zalalutdinov2003frequency} for significant applications in micro-scale and nano-scale systems \cite{RevModPhys.74.347}. For example,  synchronization of two anharmonic nanomechanical oscillators  is implemented in \cite{matheny2014phase}. And, experimentally, the synchronization measure of the system has been realized through optomechanical devices, including the synchronization of two nanomechanical beam oscillators coupled by a mechanical element \cite{shim2007synchronized}, two dissimilar silicon nitride micromechanical oscillators coupled by an optical cavity radiation field \cite{PhysRevLett.109.233906}, and two nanomechanical oscillators via a photonic resonator \cite{PhysRevLett.111.213902}. These ingenious experiments fully test the theoretical prediction of the synchronization of optomechanical systems. In addition, the relationship between quantum synchronization and the collective behavior of classical systems is also widely concerned, such as quantum synchronization of van der Pol oscillators with trapped ions \cite{lee2013quantum}, and quantum-classical transition of correlations of two coupled cavities \cite{lee2013b}. Besides, the role of the environment  and the correlation between the subsystems in the system with quantum synchronization, such as entanglement and mutual information, have been discussed as the main influencing factors \cite{giorgi2012quantum,liao2019quantum,PhysRevA.100.012133}.

Another aspect of synchronization drawing
much more attention recently is the generalization of its classical concepts into the continuous-variable quantum systems, such as complete synchronization \cite{pecora1990synchronization}, phase synchronization \cite{parlitz1996experimental,ho2002phase}, lag synchronization \cite{taherion1999observability}, and generalized synchronization \cite{zheng2000generalized}. After Mari \textit{et al.} introduced the concepts of quantum complete synchronization and quantum phase synchronization \cite{PhysRevLett.111.103605},  some interesting efforts have been devoted to enhance the quantum synchronization and quantum phase synchronization by manipulating the modulation \cite{PhysRevE.91.032910,PhysRevA.95.053858}, changing the ways of coupling between two subsystems \cite{PhysRevLett.111.103605,mari2012opto,PhysRevLett.103.213603,du2017synchronization}, or introducing nonlinearity \cite{qiao2018quantum,PhysRevE.95.022204}. Furthermore, the concepts of  quantum generalized synchronization, time-delay synchronization as well as in-phase and anti-phase synchronization have also been mentioned in \cite{PhysRevE.93.062221,PhysRevA.90.053810}. However, other than the quantum complete synchronization under which the pairs of variables have the same amplitude and phase, the concept of quantum anti-synchronization corresponding to classical anti-synchronization has not been proposed yet. Moreover,  a more generalized quantum synchronization can be defined as ``the pairs of variables have the same amplitude and possess the same $\varphi$ phase shift" (hereafter referred to as quantum $\varphi$-synchronization), i.e., for $\varphi=\pi$, the pairs of variables, such as positions and momenta, will always have a $\pi$ phase difference with each other \cite{PhysRevA.90.053810}. This type of quantum $\varphi$-synchronization is called quantum anti-synchronization. Hence, one will naturally ask how to define  and measure the quantum $\varphi$-synchronization?

To shed light on this question, in this work we give the definition of quantum $\varphi$-synchronization for the continuous-variable quantum systems by combining the concept of quantum synchronization and the phenomenon of transition from in-phase to anti-phase synchronization \cite{PhysRevA.90.053810}. The paper is organized as follows. In Sec. \ref{sec2}, we first reexamine the definitions of quantum complete synchronization and phase synchronization. Based on these concepts, the definition of quantum $\varphi$-synchronization is given, by which the quantum synchronization and quantum anti-synchronization can be treated as special cases of quantum $\varphi$-synchronization. The $\varphi$-synchronization of a coupled optomechanical system with periodic modulation is studied to illustrate our theory in Sec. \ref{sec3}. In Sec. \ref{sec4}, a brief discussion and summary are given.

\section{Measure of quantum synchronization and quantum $\varphi$-synchronization}\label{sec2}
Unlike the synchronization in classical system, the complete synchronization in quantum system can not be defined straightforwardly, since the fluctuations of the variables in the two subsystems must be strict to the limits brought by the Heisenberg principle. To address this issue, Mari \textit{et al.} proposed the measure criterion of quantum complete synchronization for continuous variable (CV) systems \cite{PhysRevLett.111.103605}
\begin{equation}
\begin{aligned}
S_{c}&=\frac{1}{\langle q_{-}(t)^{2}+ p_{-}(t)^{2}\rangle},
\label{eq1}
\end{aligned}
\end{equation}
where $q_{-}(t)=\frac{1}{\sqrt{2}}[q_{1}(t)-q_{2}(t)]$ and $p_{-}(t)=\frac{1}{\sqrt{2}}[p_{1}(t)-p_{2}(t)]$ are error operators. In order to study purely quantum mechanical effects,  the changes of variables are generally taken as
\begin{equation}
\begin{aligned}
q_{-}(t)\rightarrow \delta q_{-}(t)&=q_{-}(t)-\langle q_{-}(t) \rangle,\\
p_{-}(t)\rightarrow \delta p_{-}(t)&=p_{-}(t)-\langle p_{-}(t) \rangle.
\label{eq2}
\end{aligned}
\end{equation}
Then the contribution of the classical systematic error brought by the mean values $\langle q_{-}(t) \rangle$ and $\langle p_{-}(t) \rangle$ in $S_c$ can be dropped, and  $S_c$ will be replaced by the pure quantum synchronization measure
\begin{equation}
\begin{aligned}
S_{q}=\frac{1}{\langle \delta q_{-}(t)^{2}+ \delta p_{-}(t)^{2}\rangle}.
\label{eq3}
\end{aligned}
\end{equation}
This definition requires that the mean values of $q_{-}(t) $ and $ p_{-}(t)$ are exactly zero, i.e., $\langle q_{-}(t) \rangle =0$ and $\langle p_{-}(t) \rangle=0 $.

Mari \textit{et al.} have explained that if the averaged phase-space trajectories (limit cycles) of the two systems are constant but slightly different from
each other, a classical systematic error can be easily excluded from the measure of synchronization \cite{PhysRevLett.111.103605}. The mean-value synchronization is also regarded as a necessary condition of pure quantum synchronization \cite{PhysRevE.95.022204}. It is more reasonable and rigorous to study pure quantum synchronization based on mean-value synchronization. Therefore, we can generalize the definition of quantum complete synchronization into the quantum $\varphi$-synchronization
\begin{equation}
\begin{aligned}
S_{\varphi}&=\frac{1}{\langle q^\varphi_{-}(t)^{2}+p^\varphi_{-}(t)^{2}\rangle},
\label{eq4}
\end{aligned}
\end{equation}
which does not require mean-value synchronization. The $\varphi$-error operators are defined as $q^\varphi_{-}(t)=\frac{1}{\sqrt{2}}[q^\varphi_{1}(t)-q^\varphi_{2}(t)]$ and $p^\varphi_{-}(t)=\frac{1}{\sqrt{2}}[p^\varphi_{1}(t)-p^\varphi_{2}(t)]$ with
\begin{equation}
\begin{aligned}
q^\varphi_{j}(t)&=q_{j}(t)\cos(\varphi_{j})+p_{j}(t)\sin(\varphi_{j}),\\
p^\varphi_{j}(t)&=p_{j}(t)\cos(\varphi_{j})-q_{j}(t)\sin(\varphi_{j}),
\label{eq5}
\end{aligned}
\end{equation}
where the phase $\varphi_{j}=\arctan[\langle p_{j}(t)\rangle/\langle q_{j}(t)\rangle]$, $\varphi_{j}\in [0,2\pi]$.  The upper limit of $S_{\varphi}$ is also given by the Heisenberg principle
\begin{equation}
\begin{aligned}
S_{\varphi}&=\frac{1}{\langle q^\varphi_{-}(t)^{2}+p^\varphi_{-}(t)^{2}\rangle }\\
&\leq \frac{1}{2\sqrt{\langle q_{-}^\varphi(t)^{2} \rangle \langle p_{-}^\varphi(t)^{2} \rangle }}\\
&\leq \frac{1}{2\sqrt{[\langle q_{-}^\varphi(t)^{2} \rangle -\langle q^\varphi_{-}(t)\rangle^{2}][\langle p_{-}^\varphi(t)^{2} \rangle -\langle p^\varphi_{-}(t)\rangle^{2}]}}\\
&\leq  \frac{1}{\sqrt{\mid  \frac{1}{2}[q^\varphi_{1}(t),p^\varphi_{1}(t)]+\frac{1}{2}[q^\varphi_{2}(t),p^\varphi_{2}(t)]\mid^{2}}}\\
&=1.
\label{eq6}
\end{aligned}
\end{equation}
This means that the closer $S_{\varphi}$ is to 1, the better the quantum $\varphi$-synchronization. Again, let's take the changes of variables
\begin{equation}
\begin{aligned}
&q^\varphi_{-}(t)\rightarrow \delta q^\varphi_{-}(t)=q^\varphi_{-}(t)-\langle q^\varphi_{-}(t) \rangle,\\
&p^\varphi_{-}(t)\rightarrow \delta p^\varphi_{-}(t)=p^\varphi_{-}(t)-\langle p^\varphi_{-}(t) \rangle.
\label{eq7}
\end{aligned}
\end{equation}
The mean values of $q^\varphi_{-}(t) $ and $ p^\varphi_{-}(t)$ are zero when the average amplitude and period of the mean value of the two variables are the same.
In this case, $S_{\varphi}$ equals to the pure quantum $\varphi$-synchronization measure $S^{\varphi}_{q}$ mathematically
\begin{equation}
\begin{aligned}
S^{\varphi}_{q}&=\frac{1}{\langle \delta q^\varphi_{-}(t)^{2}+\delta p^\varphi_{-}(t)^{2}\rangle }\\
&= \left\langle\frac{1}{2}\left[(\delta p_1)^2+(\delta q_1)^2+(\delta p_2)^2+(\delta q_2)^2\right.\right.\\
&+2(\delta p_1\delta q_2-\delta q_1\delta p_2)\sin{\varphi}
-2(\delta p_1\delta p_2+\delta q_1\delta q_2)\cos{\varphi}]\rangle^{-1},
\label{eq8}
\end{aligned}
\end{equation}
where $\varphi=\varphi_{2}-\varphi_{1}$ can be determined by the final steady state. Now let's explain the relationship between quantum $\varphi$-synchronization and quantum complete synchronization, quantum phase synchronization. We can see from Eq. (\ref{eq8}) that the definition of $S^{\varphi}_{q}$ can be reduced into (a) quantum synchronization: if $\varphi=0$, then $S_{q}=S_q^{\varphi}$; (b) quantum phase synchronization: if $\langle \delta q^\varphi_{-}(t)^{2}\rangle=\langle\delta p^\varphi_{-}(t)^{2}\rangle$, then $S_{p}=S_q^{\varphi}$. (c) if $\varphi=\pi$, $\widetilde{S}_{q}=S_q^{\varphi}$, where $\widetilde{S}_{q}$ can be defined as quantum anti-synchronization. Therefore, quantum synchronization and quantum anti-synchronization are the special cases of quantum $\varphi$-synchronization. But the definition of quantum phase synchronization  is slightly different \cite{PhysRevLett.111.103605}
\begin{equation}
S_p=\frac{1}{2}\langle \delta p^{\varphi}_{-}(t)^2\rangle^{-1}
=\frac{1}{\langle\delta p^{\varphi}_{-}(t)^2 +\delta p^{\varphi}_{-}(t)^2\rangle}.
\label{eq9}
\end{equation}
Unlike $S_\varphi$ in Eq. (\ref{eq6}), the measure of quantum phase synchronization $ S_p$ can exceed $1$. To illustrate these definitions,  we next compare quantum $\varphi$-synchronization with quantum synchronization and quantum phase synchronization in coupled optomechanical system with periodic modulation.
\section{Quantum synchronization, quantum phase synchronization and quantum $\varphi$-synchronization in coupled optomechanical system with periodic modulation}\label{sec3}
\begin{figure}[t]
\centering
\includegraphics*[width=0.9\columnwidth]{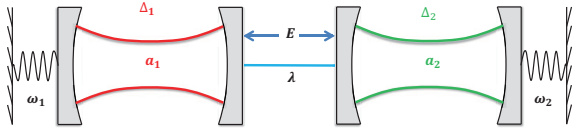}
\caption{Schematic illustration of coupled optomechanical system with periodic modulation.}
\label{fig1}
\end{figure}

To  examine the relations and differences between quantum synchronization, quantum phase synchronization and quantum $\varphi$-synchronization, we consider a coupled optomechanical system with periodic modulation \cite{du2017synchronization,qiao2018quantum}. Two subsystems are coupled by optical fibers \cite{PhysRevA.85.033805}. Each of them is consisted of a mechanical oscillator coupled with a Fabry-Perot cavity driven by a time-periodic modulated filed (see Fig. \ref{fig1}) \cite{PhysRevA.96.023805}. It's worth noting that the optomechanical device is experimentally possible. On the one hand, the synchronization of two mechanically isolated nanomechanical resonators via a photonic resonator has been implemented \cite{PhysRevLett.111.213902}. On the other hand, the self-oscillating mechanical resonator in on-fiber optomechanical cavity excited by a tunable laser with periodically modulated power also has been studied \cite{PhysRevE.91.032910}. Then the Hamiltonian of the whole coupled system can be written as
\begin{equation}
\begin{aligned}
H=&\sum_{j=1}^{2}\left\{-\triangle_{j}\left[1+A_{c}\cos(\omega_{c} t)\right]a_{j}^{\dagger}a_{j}+\frac{\omega_{j}}{2}\left(p_{j}^{2}+q_{j}^{2}\right)-ga_{j}^{\dagger}a_{j}q_{j}\right.\\
   &\left.+iE(a_{j}^{\dagger}-a_{j})\right\}+\lambda\left(a_{1}^{\dagger}a_{2}+ a_{2}^{\dagger}a_{1}\right),
\label{eq10}
\end{aligned}
\end{equation}
where $a$ and $a^{\dagger}$ are the creation and annihilation operators, $q_{j}$ and $p_{j}$ are the position and momentum operators of mechanical oscillator with frequency $\omega_{j}$ in $j$th subsystem , respectively \cite{wang2016steady,jin2017reconfigurable}. $\lambda$ is the optical coupling strength and $E$ is the intensity of the driving field. $\Delta_{j}$ is the optical detuning, which is modulated with a common frequency $\omega_c$ and amplitude $A_c$. $g$ is the optomechanical coupling constant.
To solve time evolution of the dynamical operators $O=q_j,p_j,a_j$ of the system, we consider the dissipation effects in the Heisenberg picture and utilize the quantum Langevin equation \cite{aspelmeyer2014cavity}. From Eq. (\ref{eq10}), the evolution equations of the operators can be written as:
\begin{equation}
\begin{aligned}
\dot{q_{j}}=&\omega_{j}p_{j}, \\
 \dot{p_{j}}=&-\omega_{j}q_{j}-
 \gamma p_{j}+ga_{j}^{\dagger}a_{j}+\xi_{j}, \\
\dot{a_{j}}=&-\left\{\kappa-i\triangle_{j}\left[1+A_{c}\cos(\omega_{c}t)\right]\right\}a_{j}+iga_{j}q_{j}+E\\
&-i\lambda a_{3-j}+\sqrt{2\kappa}a_{j}^{in},
 \label{eq11}
\end{aligned}
\end{equation}
where $\kappa$ is the radiation loss coefficient \cite{PhysRevLett.113.053604,schonleber2016optomechanical} and $\gamma$ is mechanical damping rate, respectively. $a_{j}^{in}$ and $\xi_{j}$ are input bath operators and satisfy standard correlation: $\left\langle a^{in\dagger}(t)a^{in}(t')+a^{in}(t')a^{in\dagger}(t)\right\rangle=\delta(t-t')$ and  $\frac{1}{2}\langle \xi_{j}(t)\xi_{j'}(t')+\xi_{j'}(t')\xi_{j}(t)\rangle=\gamma(2\bar{n}_{bath}+1)\delta_{jj'}\delta(t-t')$ under the Markovian approximation \cite{wang2016steady,jin2017reconfigurable}, where $\bar{n}_{bath}=1/[\exp\left(\hbar \omega_{j}/k_{B}T\right)-1]$ is the mean occupation number of the mechanical baths which gauges the temperature $T$ of the system \cite{giovannetti2001phase,liu2014optimal,xu2015optical}. To solve the set of nonlinear differential operator equations , we need to linearize Eq. (\ref{eq11}). There may be several ways to do that, such as using the stochastic Hamiltonian \cite{PhysRevA.94.031802,PhysRevLett.118.233604}, and mean field approximation\cite{li2015criterion,PhysRevE.93.062221,farace2012enhancing,mari2012opto}. Here we use the mean field approximation, since it can uncover the effects of mean error and quantum fluctuation on quantum synchronization. Namely, the operators are decomposed into an mean value and a small fluctuation, i.e.
\begin{equation}
  O(t)=\langle O (t)\rangle+\delta O(t).
\label{eq12}
\end{equation}
And, as long as $|\langle O (t)\rangle|\gg 1$, the usual linearization approximation to Eq. (\ref{eq11}) can be implemented \cite{PhysRevLett.103.213603}.
Then, Eq. (\ref{eq11}) can be divided into two different sets of equations, one for the mean value
\begin{equation}
\begin{aligned}
\partial_{t}{\langle q_{j}\rangle}=& \omega_{j}\langle p_{j}\rangle , \\
\partial_{t}{\langle p_{j}\rangle}=& -\omega_{j}\langle q_{j}\rangle-\gamma \langle p_{j}\rangle +g|\langle a_{j}\rangle |^{2}, \\
\partial_{t}{\langle a_{j}\rangle}=&-\left\{\kappa-i\triangle_{j}\left[1+A_{c}\cos(\omega_{c}t)\right]\right\}\langle a_{j} \rangle +ig\langle a_{j}\rangle\langle q_{j}\rangle+E\\
&-i\lambda \langle a_{3-j} \rangle ,
\label{eq13}
\end{aligned}
\end{equation}
and the other for the fluctuation:
\begin{equation}
\begin{aligned}
\partial_{t}{\delta q_{j}}=& \omega_{j}\delta p_{j}, \\
\partial_{t}{\delta p_{j}}=& -\omega_{j}\delta q_{j}-\gamma\delta p_{j}+g(\langle a_{j}\rangle \delta a_{j}^{\dagger}+\langle a_{j}\rangle^{*}\delta a_{j}) +\xi_{j},\\
\partial_{t}{\delta a_{j}}=&-\left\{\kappa-i\triangle_{j}\left[1+A_{c}\cos(\omega_{c}t)\right]\right\}\delta a_{j}+ig(\langle a_{j}\rangle \delta q_{j}\\&
+\langle q_{j}\rangle \delta a_{j})-i\lambda \delta a_{3-j}+\sqrt{2\kappa}a_{j}^{in}.
\end{aligned}
\label{eq14}
\end{equation}
In Eq. (\ref{eq14}), the second-order smaller terms of the fluctuation have been ignored. Then, by defining $u=(\delta q_{1}, \delta p_{1}, \delta x_{1}, \delta y_{1}, \delta q_{2}, \delta p_{2}, \delta x_{2}, \delta y_{2})^{\top} $ with $\delta x_{j}=\frac{1}{\sqrt{2}}\left(\delta a_{j}^{\dagger}+\delta a_{j}\right)$ and $\delta y_{j}=\frac{i}{\sqrt{2}}\left(\delta a_{j}^{\dagger}-\delta a_{j}\right)$, Eq. (\ref{eq14}) can be simplified to:
\begin{equation}
\partial_{t}{u}=Mu+n,
\label{eq15}
\end{equation}
where $n=(0, \xi_{1}, \sqrt{2\kappa}x_{1}^{in}, \sqrt{2\kappa}y_{1}^{in}, 0, \xi_{2}, \sqrt{2\kappa}x_{2}^{in}, \sqrt{2\kappa}y_{2}^{in})^{\top}$ is the noise vector with  $x_{1}^{in}=\frac{1}{\sqrt{2}}\left(a^{in^{\dagger}}+a^{in}\right)$ and $y_{1}^{in}=\frac{i}{\sqrt{2}}\left(a^{in^{\dagger}}-a^{in}\right)$. $M$ is a time-dependent coefficient matrix:
\begin{equation}
M=\left(\begin{array}{cc}
M_{1} &M_{0}\\
M_{0} &M_{2}\\
\end{array}\right),
\label{eq16}
\end{equation}
with
$$M_{j}=\left(\begin{array}{cccc}
0&\omega_{j} & 0&0\\
-\omega_{j} &-\gamma &\sqrt{2}g \mathrm{Re}(\langle a_j\rangle)&\sqrt{2}g \mathrm{Im}(\langle a_j\rangle)\\
-\sqrt{2}g \mathrm{Im}(\langle a_j\rangle)&0&-\kappa&-F_{j}\\
\sqrt{2}g \mathrm{Re}(\langle a_j\rangle)&0&F_{j} &-\kappa
\end{array}\right),$$
and
$$M_{0}=\left(\begin{array}{cccc}
0&0& 0&0 \\
0 &0&0&0\\
0 &0&0&\lambda\\
0 &0&-\lambda&0
\end{array}\right),$$
where $F_{j}=\Delta_{j}[1+A_{c}\cos(\omega_{c}t)]+g\langle q_{j}\rangle$ and the evolution process of matrix element $M(t)$ at any time can be obtained by solving Eq. (\ref{eq13}) numerically when the initial conditions are $0$. In order to study the contribution of quantum fluctuation to quantum synchronization, we consider the following covariance matrix:
\begin{equation}
V_{ij}\equiv\frac{1}{2}\langle u_{i}u_{j}+u_{j}u_{i}\rangle.
\label{eq17}
\end{equation}
The evolution of $V$ over time is governed by \cite{li2015criterion,PhysRevLett.103.213603,wang2014nonlinear,larson2011photonic}:
\begin{equation}
\partial_{t}{V}=MV+VM^{T}+N.
\label{eq18}
\end{equation}
\begin{figure}[t]
\centering
\includegraphics[width=0.95\columnwidth]{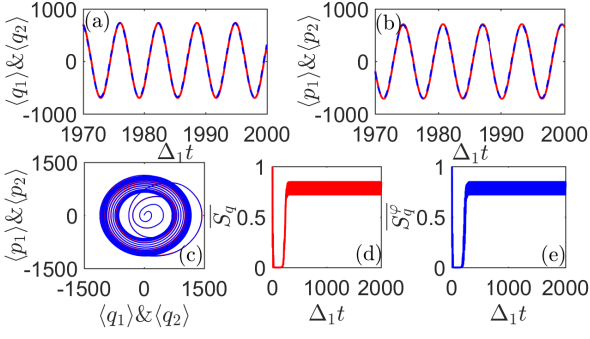}
\caption{Time evolution of (a)  the mean values $\langle q_{1}\rangle$ (red solid line) and $\langle q_{2}\rangle$ (blue dashed line) , (b)  the mean values $\langle p_{1}\rangle$ (red solid line) and $\langle p_{2}\rangle$ (blue dashed line), (c) the limit-cycle trajectories in the $\langle q_{1}\rangle\leftrightharpoons\langle p_{1}\rangle$ (red) and $\langle q_{2}\rangle\leftrightharpoons\langle p_{2}\rangle$ (blue) spaces, (d) the measure of quantum synchronization $\overline{S_{q}}$, and (e) the measure of quantum $\varphi$-synchronization $\overline{S^{\varphi}_{q}}$. Parameters are chosen to refer to \cite{PhysRevLett.111.103605,PhysRevA.85.033805,PhysRevE.95.022204,du2017synchronization}: $\lambda=0.03\Delta_{1}$, $A_{c}=2,\omega_{c}=3\Delta_{1}$. Other parameters are normalized by $\Delta_{1}=1$,$\Delta_{2}=1.005\Delta_{1}$,$\omega_{1}=\Delta_{1}$,$\omega_{2}=\Delta_{2}$, $g=0.005 \Delta_{1}$, $\gamma=0.005\Delta_{1}$, $\kappa=0.15\Delta_{1}$, $E=100 \Delta_{1}$.}
\label{fig2}
\end{figure}
The noise matrix $N=\mathrm{diag}(0,\gamma(2\bar{n}_{bath}+1),\kappa,\kappa,0,\gamma(2\bar{n}_{bath}+1),\kappa,\kappa)$ satisfying $N_{ij}\delta(t-t^{'})=\frac{1}{2}\langle \xi_{i}(t)\xi_{j}(t^{'})+\xi_{j}(t^{'})\xi_{i}(t) \rangle$. Hence, Eq. (\ref{eq3}), Eq. (\ref{eq8}) and Eq. (\ref{eq9}) can be rewritten in terms of $V_{ij}$
\begin{equation}
\begin{aligned}
S_{q}=&\left[\frac{1}{2}\left( V_{11}+V_{22}+V_{55}+V_{66}-V_{15}-V_{51}-V_{62}-V_{26} \right)\right]^{-1},\\
S^{\varphi}_{q}=&\left[\frac{1}{2}( V_{11}+V_{22}+V_{55}+V_{66}+2V_{25}\sin{\varphi}-2V_{16}\sin{\varphi}\right.\\
&-2V_{26}\cos{\varphi}-2V_{15}\cos{\varphi})]^{-1},\\
S_p=& [V_{11}(\sin{\varphi_1})^2+V_{22}(\cos{\varphi_1})^2+V_{55}(\sin{\varphi_2})^2\\
&+V_{66}(\cos{\varphi_2})^2-2V_{12}\sin{\varphi_1}\cos{\varphi_1}-2V_{15}\sin{\varphi_1}\sin{\varphi_2}\\
&+2V_{16}\sin{\varphi_1}\cos{\varphi_2}+2V_{25}\cos{\varphi_1}\sin{\varphi_2}\\
&-2V_{26}\cos{\varphi_1}\cos{\varphi_2}-2V_{56}\cos{\varphi_2}\sin{\varphi_2}
]^{-1},
\label{eq19}
\end{aligned}
\end{equation}
and their evolutions can be derived by solving Eq. (\ref{eq13}), Eq. (\ref{eq15}) and Eq. (\ref{eq18}). In addition, under different parameters, the calculated time-averaged synchronization
\begin{equation}
\begin{aligned}
\overline{S_o}(t)=\lim_{T\rightarrow \infty}\frac{1}{T} \int_{0}^{T} S_o(t) dt
\end{aligned}
\end{equation}
is used as the synchronization measure in the asymptotic steady state of the system, where $o=\varphi,p$. According to $R-H$ criterion \cite{PhysRevA.35.5288}, all the eigenvalues of coefficient matrix $M$ will be negative after a temporary evolutionary process. Hence, a stable limit cycle solution representing a periodic oscillation will exit \cite{geng2018enhancement}.\\

\begin{figure}[t]
\centering
\includegraphics[width=0.95\columnwidth]{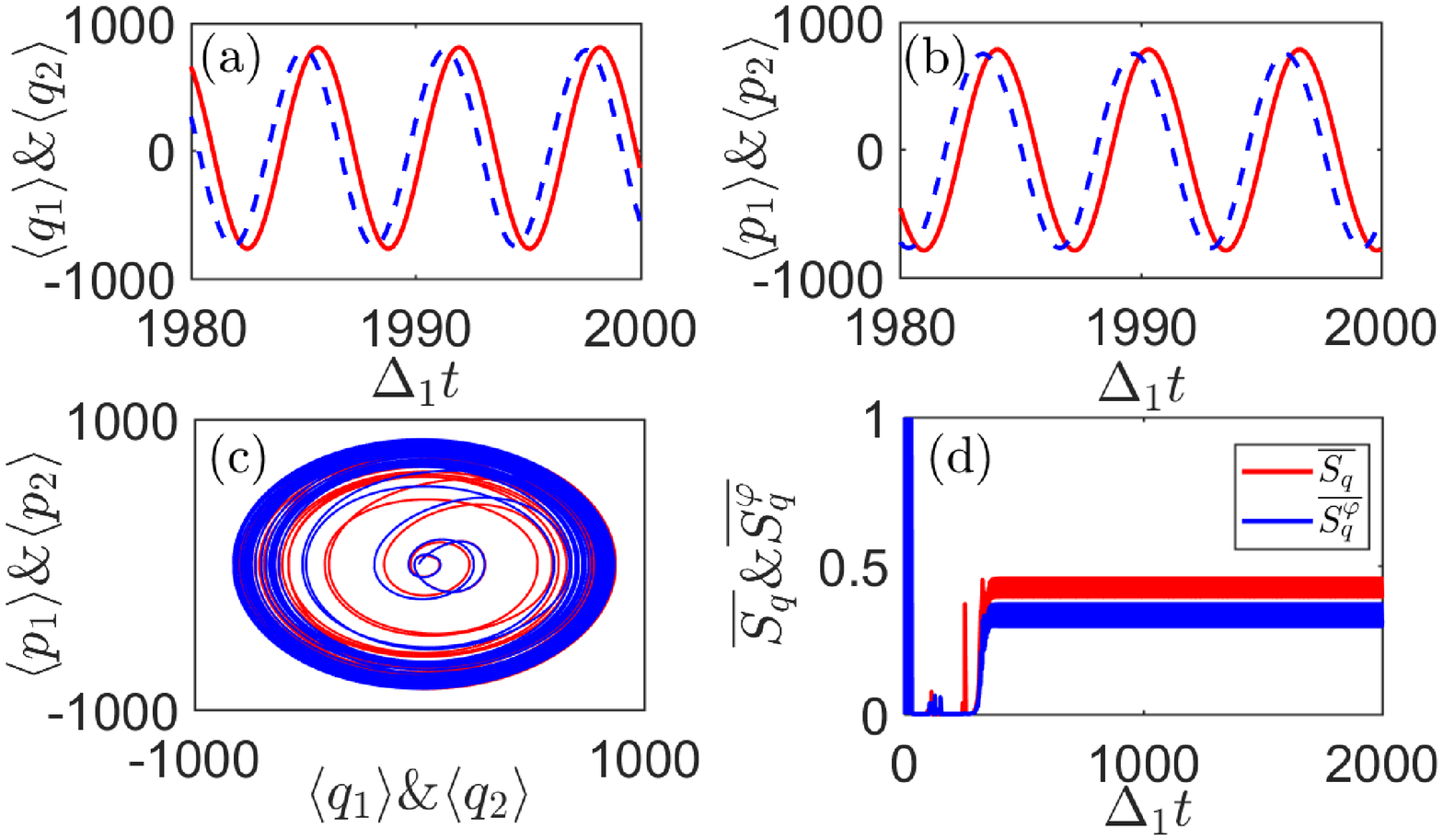}
\caption{Time evolution of (a) the mean values $\langle q_{1}\rangle$ (red solid line) and $\langle q_{2}\rangle$ (blue dashed line), (b) the mean value $\langle p_{1}\rangle$ (red solid line) and $\langle p_{2}\rangle$(blue dashed line), (c) the limit-cycle trajectories in the $\langle q_{1}\rangle\leftrightharpoons\langle p_{1}\rangle$ (red) and $\langle q_{2}\rangle\leftrightharpoons\langle p_{2}\rangle$ (blue) spaces, and (d) time evolution of the measure of quantum synchronization $\overline{S_{q}}$ [the red (upper) line] and quantum $\varphi$-synchronization $\overline{S^{\varphi}_{q}}$ [the blue (lower) line] with $\lambda=0.14\Delta_{1}$ and $A_{c}=1,\omega_{c}=2\Delta_{1}$. The other parameters are the same as in Fig. \ref{fig2}.}.
\label{fig3}
\end{figure}

\begin{figure}[t]
\centering
\includegraphics*[width=0.9\columnwidth]{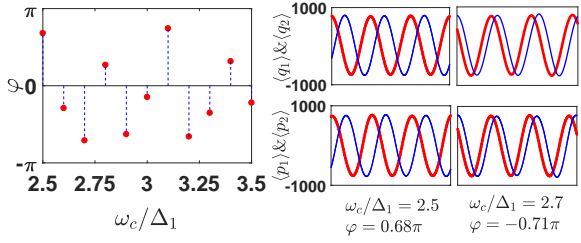}
\caption{Discrete point diagram (left) of the phase $\varphi$ versus modulation frequency $\omega_c$ when the system reaches steady state. The time evolution (right) of the mean values $\langle q_{1}\rangle\&\langle p_{1}\rangle$ (red thick line) and $\langle q_{2}\rangle\& \langle p_{2}\rangle$ (blue thin line) with two values of $\omega_c$ from the left panel. The parameter $\lambda=0.14\Delta_{1}$, $A_{c}=1$ and other parameters are the same as in Fig. \ref{fig2}.}
\label{fig6}
\end{figure}

As we discussed in the last section, if $\varphi=0$, which requires the condition of mean-value complete synchronization, i.e., $\langle q_{-}(t) \rangle =\langle q_{1}(t)\rangle -\langle q_{2}(t)\rangle=0$ , $\langle p_{-}(t) \rangle=\langle p_{1}(t)\rangle -\langle p_{2}(t)\rangle=0$, the measures of quantum $\varphi$-synchronization and quantum synchronization are equivalent, i.e., $S^{\varphi}_{q}=S_{q}$. As shown in Fig. \ref{fig2}(a), $\langle q_{1}(t)\rangle$ and $\langle q_{2}(t)\rangle$ are found to oscillate exactly in phase when entering the stable state. The same conclusion holds for $\langle p_{1}(t)\rangle$ and $\langle p_{2}(t)\rangle$ in Fig. \ref{fig2}(b). In Fig. \ref{fig2}(c), the evolutions of $\langle q_{1}(t) \rangle \leftrightharpoons\langle p_{1}(t) \rangle$ and $\langle q_{2}(t) \rangle \leftrightharpoons\langle p_{2}(t) \rangle$ of the two oscillators trend to an asymptotic periodic orbit (i.e. the two limit cycles tend to be consistent), which indicates that the system is stable. And, Fig \ref{fig2}(d) and (e) show that the changes in $\overline{S_q}$ and $\overline{S^{\varphi}_{q}}$ over time are exactly the same.
When the mean-value synchronization is not
complete, as shown in Fig. \ref{fig3}(a) and \ref{fig3}(b), there exists a phase advance between $\varphi_2$ and $\varphi_1$, i.e $\varphi=\varphi_2-\varphi_1=\arctan[\langle p_{2}(t)\rangle/\langle q_{2}(t)\rangle]-\arctan[\langle p_{1}(t)\rangle/\langle q_{1}(t)\rangle]\approx0.2\pi$. Similarly, the two consistent limit cycles are shown  in Fig. \ref{fig3}(c), indicating that the evolution of the system can still reach a steady state when  mean-value is not complete synchronization.
However, quantum synchronization $\overline{S_q}$ and quantum $\varphi$-synchronization are different as shown in Fig. \ref{fig3}(d). This is because the definition of quantum $\varphi$-synchronization takes the effect of mean-value incomplete synchronization into account. The quantum $\varphi$-synchronization is then more general and rigorous than quantum synchronization. As the mean-value incomplete synchronization will break the condition for Eq. (\ref{eq2}), the contribution of $\langle q_{-}(t) \rangle$ and $\langle p_{-}(t) \rangle$ to the quantum complete synchronization $S_c$ is much greater than that of the quantum fluctuation. Besides, the mean-value incomplete synchronization will always occur with the change of parameters. As shown in Fig. \ref{fig6}, different phase differences $\varphi$ will be generated by the different modulation frequency $\omega_c$. And similar phenomena have been shown in \cite{PhysRevA.90.053810}. Therefore, it is necessary to give the quantum synchronization when mean-value synchronization is not complete, namely quantum $\varphi$-synchronization.

\begin{figure}[b]
\centering
\subfigure{
\label{4a}
\includegraphics[width=0.48\columnwidth]{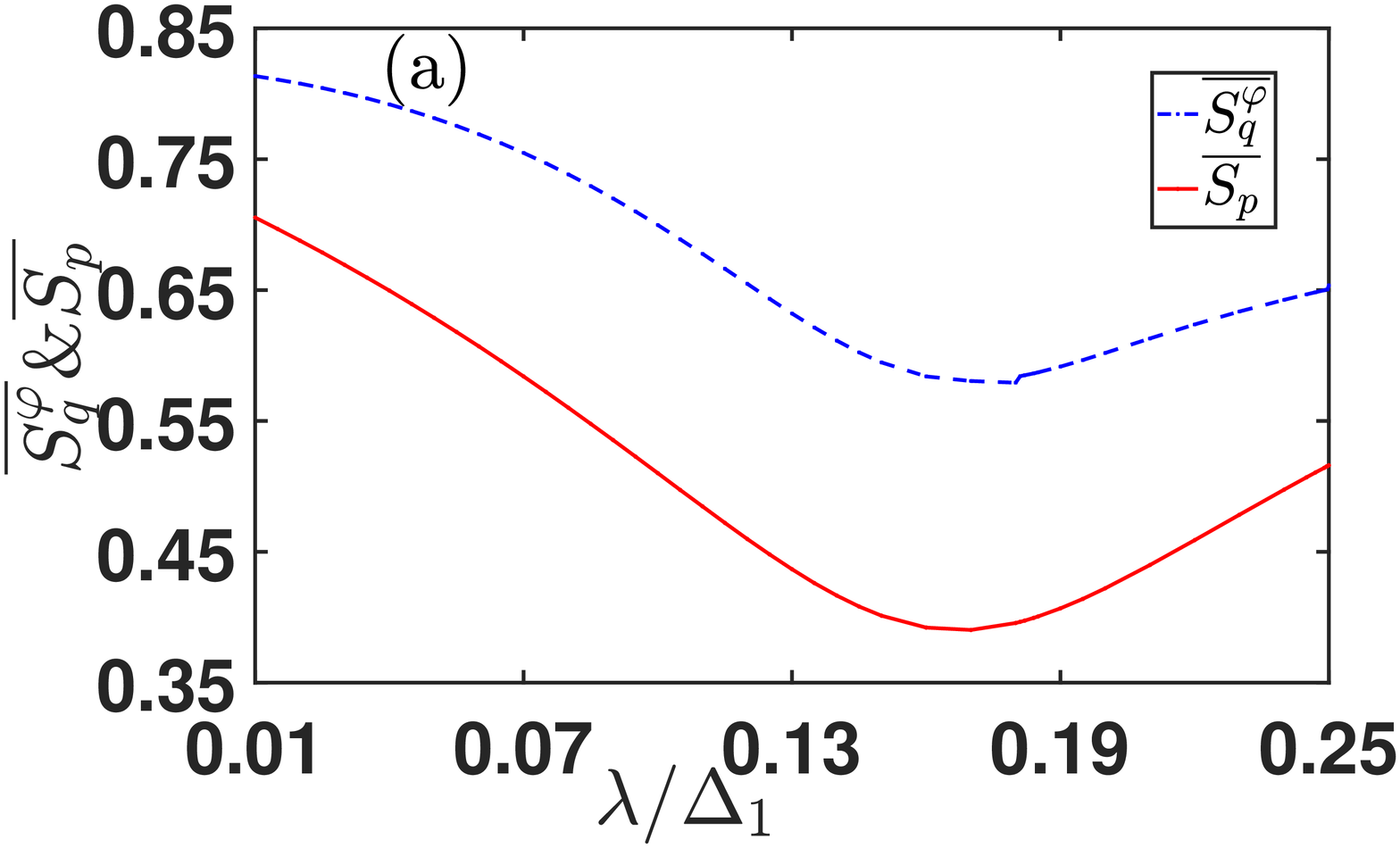}}
\centering
\subfigure{
\label{4b}
\includegraphics[width=0.48\columnwidth]{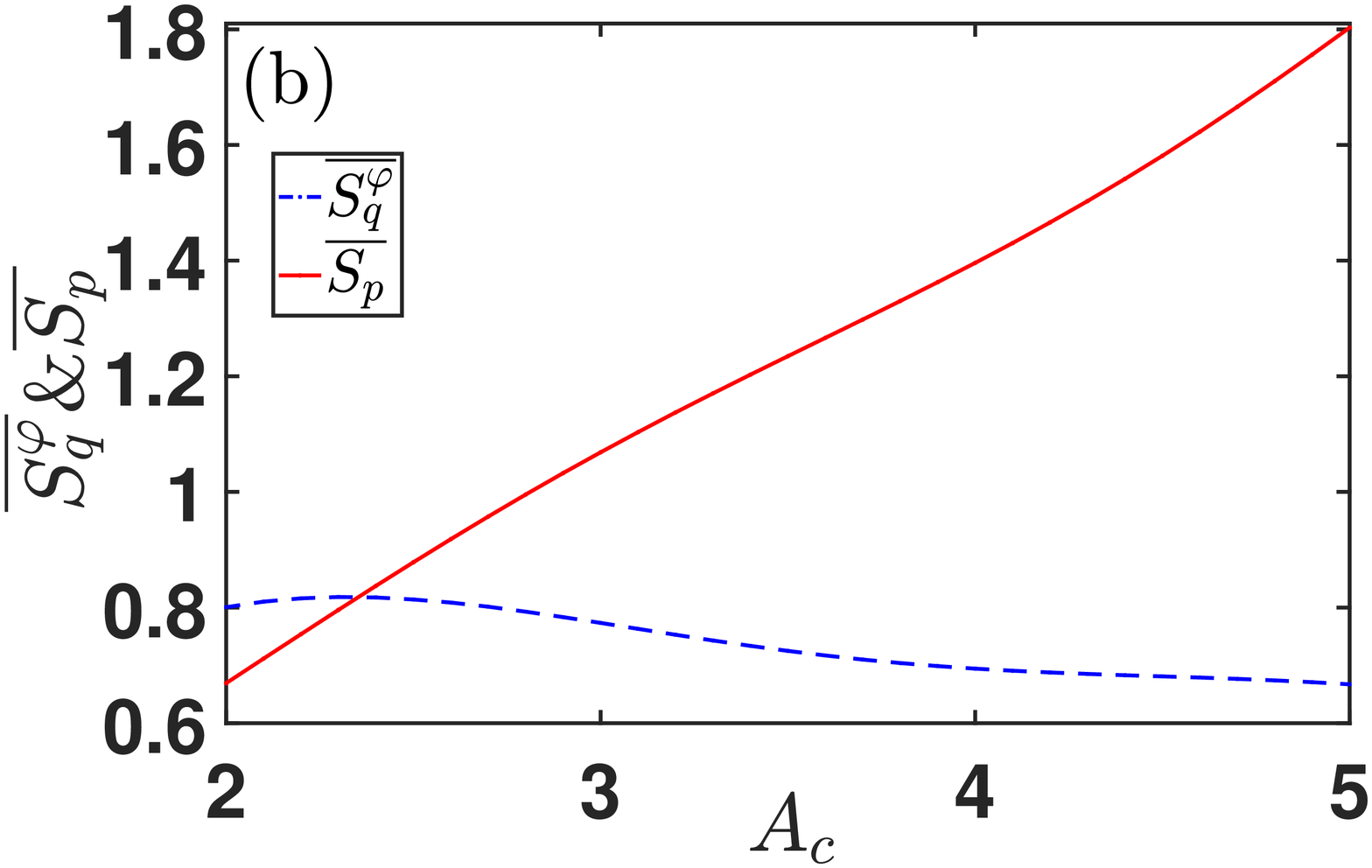}}
\caption{Mean values of the quantum phase synchronization measure $\overline{S_p}$ (red solid line)  and quantum $\varphi$-synchronization measure $\overline{S^{\varphi}_{q}}$ (the blue dotted line)
as a function of (a) the optical coupling coefficient $\lambda$ with $A_{c}=2$, $\omega_{c}=3\Delta_{1}$,
 (b) the modulation frequency $A_{c}$ with $\lambda=0.03\Delta_{1}$, $\omega_{c}=3\Delta_{1}$.
The other parameters are the same as in Fig. \ref{fig2}.}
\label{fig4}
\end{figure}
Moreover, the quantum $\varphi$-synchronization also can be related to the quantum phase synchronization. As shown in Fig. \ref{fig4}(a), both quantum $\varphi$-synchronization $\overline{S_{q}^{\varphi}}$ and quantum phase synchronization $\overline{S_p}$ first decrease and then increases as the optical coupling strength $\lambda$ increases, and the changing trend of $\overline{S_p}$ and $\overline{S^{\varphi}_{q}}$ with $\lambda$ is accordant. When $\lambda=0.016\Delta_{1}$, both $\overline{S_{\varphi}}=0.58$ and $\overline{S_p}=0.36$ are minimized. This means that $\langle \delta q_{-}^{\varphi}(t)^2 \rangle$ is approximately proportional to $\langle \delta p_{-}^{\varphi}(t)^2 \rangle$  ($\langle \delta q_{-}^{\varphi}(t)^2 \rangle>\langle \delta p_{-}^{\varphi}(t)^2 \rangle$). In this case, the measure of $\varphi$-synchronization is accordance with that of the phase synchronization. When  $\langle \delta q_{-}^{\varphi}(t)^2 \rangle=\langle \delta p_{-}^{\varphi}(t)^2 \rangle$, the two definitions are the same.  However, if  $\langle \delta q_{-}^{\varphi}(t)^2 \rangle$ has no linear relation with $\langle \delta p_{-}^{\varphi}(t)^2 \rangle$, the definitions of $\varphi$-synchronization and phase synchronization are quite different as shown in
Fig. \ref{fig4}(b). In Fig. \ref{fig4}(b), the quantum $\varphi$-synchronization $\overline{S^{\varphi}_q}$ becomes worse when the modulation amplitude $A_c$ increases. While the quantum phase synchronization $\overline{S_p}$ is significantly enhanced. This difference is due to the fact that the quantum $\varphi$-synchronization takes both of $\langle \delta q_{-}^{\varphi}(t)^2 \rangle$ and $\langle \delta p_{-}^{\varphi}(t)^2 \rangle$ into consideration, while the quantum phase synchronization only considers  $\langle \delta p_{-}^{\varphi}(t)^2 \rangle$. This also results in quantum phase synchronization $\overline{S_p}$ exceeding 1 as shown in Fig. \ref{fig4}(b). However quantum $\varphi$-synchronization is still less than $1$ due to the Heisenberg principle, which has also been demonstrated in Eq. (\ref{eq6}).
\begin{figure}[t]
\centering
\subfigure{
\label{5a}
\includegraphics[width=0.75\columnwidth]{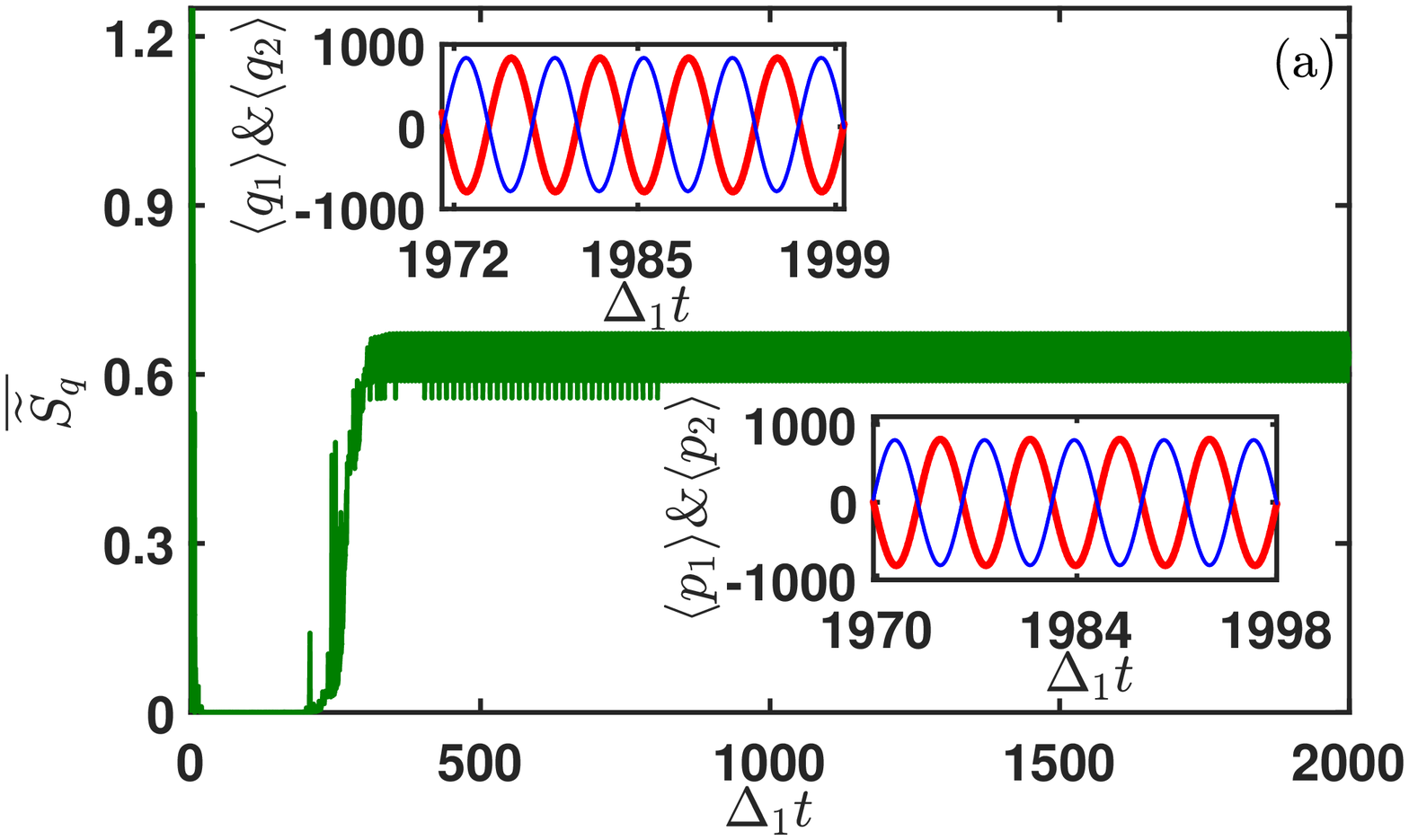}}
\centering
\subfigure{
\label{5b}
\includegraphics[width=0.75\columnwidth]{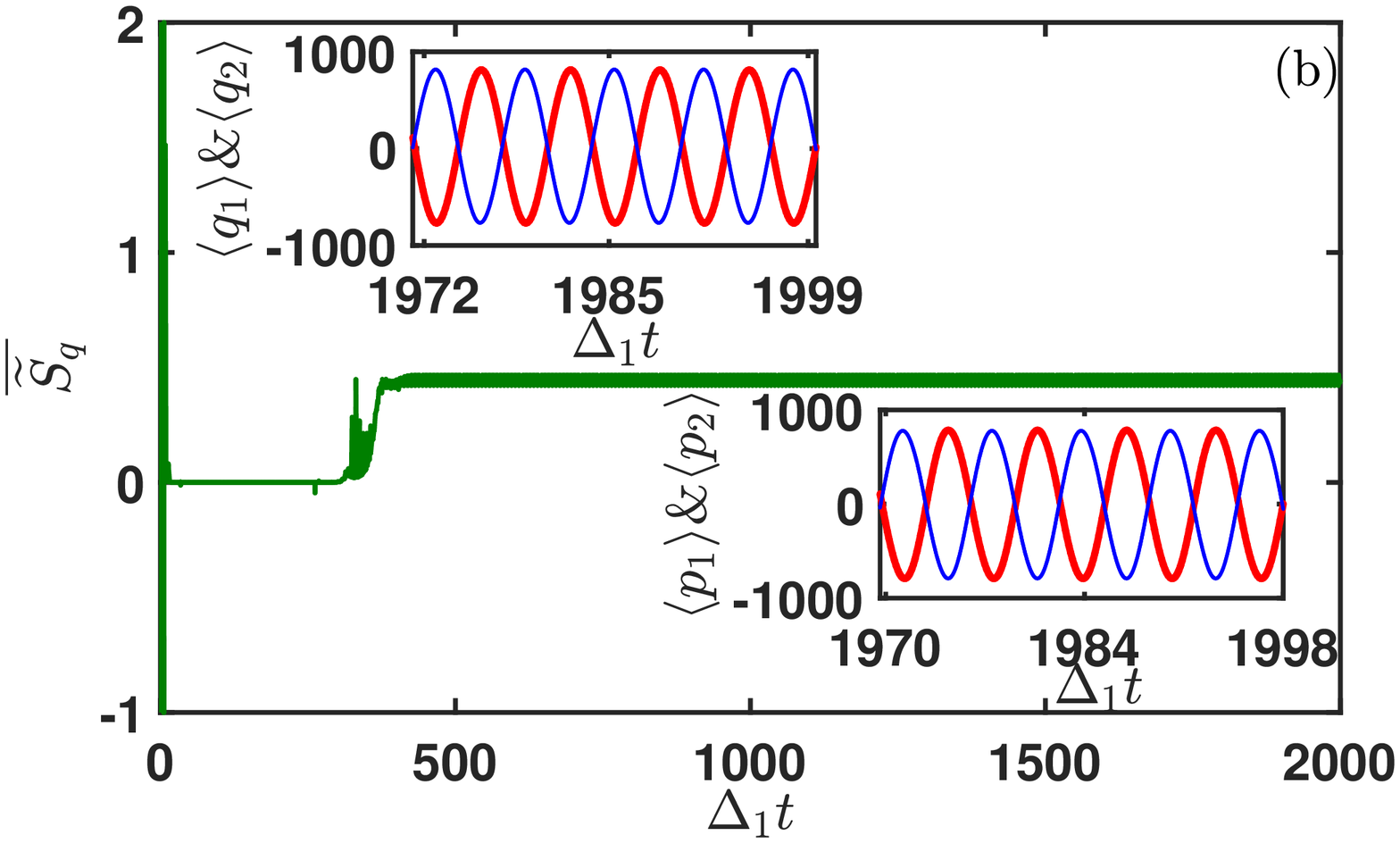}}
\caption{The evolution of the quantum synchronization $\overline{\widetilde{S}_q}$ (green solid line), and the mean value $\langle q_{1}\rangle\&\langle p_{1}\rangle$ (red thick line), $\langle q_{2}\rangle\&\langle p_{2}\rangle$ (blue thin line) when the system is stable with (a) $\lambda=0.3\Delta_{1}$ and $A_c=1.5,\omega_c=2\Delta_{1}$, (b) $\lambda=0.2\Delta_{1}$ and $A_c=1,\omega_c=2\Delta_{1}$. The other parameters are the same as in Fig. \ref{fig2}.}
\label{fig5}
\end{figure}

When $\varphi=\pi$, the $\varphi$-error operators become $q^\pi_{-}(t)=\frac{1}{\sqrt{2}}[q_{1}(t)+q_{2}(t)]$ and $p^\pi_{-}(t)=\frac{1}{\sqrt{2}}[p_{1}(t)+p_{2}(t)]$. The quantum $\varphi$-synchronization becomes quantum anti-synchronization, i.e.,
\begin{equation}
\begin{aligned}
\widetilde{S}_{q}\equiv S^{\pi}_{q}&=\frac{1}{\langle \delta q^\pi_{-}(t)^{2}+\delta p^\pi_{-}(t)^{2}\rangle }\\
&= \left\langle\frac{1}{2}\left[(\delta p_1+\delta p_2)^2+(\delta q_1+\delta q_2)^2\right]\right\rangle^{-1}.\\
\end{aligned}
\end{equation}
We can also find this phenomenon of quantum anti-synchronization in coupled optomechanical system under certain parameters. As shown in Fig. \ref{fig5}, quantum anti-synchronization is that the mean-value is anti-synchronization and quantum $\varphi$-synchronization is not zero.
\section{Conclusions} \label{sec4}
In summary, we have introduced and characterized a more generalized concept called quantum $\varphi$-synchronization. It can be defined as  the pairs of variables which have the same amplitude and possess same $\varphi$ phase shift.  The measure of  the quantum $\varphi$-synchronization has also been defined with the phase difference $\varphi$. Therefore, the quantum synchronization and quantum anti-synchronization can be treated as the special cases of quantum $\varphi$-synchronization. Besides, the quantum phase synchronization can also be related with the quantum $\varphi$-synchronization. As an example, we have investigated the quantum $\varphi$-synchronization and quantum phase synchronization phenomena of two coupled optomechanical systems with periodic modulation. It has been shown that quantum $\varphi$-synchronization is more general as a measure of synchronization than the quantum synchronization. We have also showed the different affections of the optical coupling coefficient and the modulation amplitude on the quantum phase synchronization and the quantum $\varphi$-synchronization. These two definitions of synchronization are only accordant with each in the case that  $\langle \delta q_{-}^{\varphi}(t)^2 \rangle$ is approximately proportional to $\langle \delta p_{-}^{\varphi}(t)^2 \rangle$. Based on quantum $\varphi$-synchronization, the quantum anti-synchronization phenomenon has also been defined and observed for $\varphi=\pi$ under some parameters. Therefore, the definition of quantum $\varphi$-synchronization provides a new way to study the quantum synchronization of continuous variable systems. In addition, it is interesting in the future to study the linearization method by using the stochastic Hamiltonian \cite{PhysRevA.94.031802,PhysRevLett.118.233604} and its influence on the quantum $\varphi$-synchronization.

\begin{acknowledgments}
We thank Y. Li and W. L. Li for helpful discussions. This work is supported by National Natural Science Foundation of China (NSFC) (Grants Nos. 11875103 and 11775048), the China Scholarship Council (Grant No. 201806625012), the Scientific and Technological Program of Jilin Educational Committee during the Thirteenth Five-year Plan Period (Grant Nos. JJKH20180009KJ and JJKH20190276KJ) and the Fundamental Research Funds for the Central Universities (Grant No. 2412019FZ040).
\end{acknowledgments}

\providecommand{\noopsort}[1]{}\providecommand{\singleletter}[1]{#1}%

\end{document}